\documentclass[%
 reprint,
superscriptaddress,
 amsmath,amssymb,
 aps,
showpacs,
]{revtex4-1}

\usepackage{graphicx}
\usepackage{dcolumn}
\usepackage{bm}
\usepackage{cleveref}
\usepackage{xcolor}
\usepackage{tabu}
\usepackage{lineno}

\DeclareMathAlphabet{\pazocal}{OMS}{zplm}{m}{n}

\newcommand{\mev}{\mathrm{MeV}}
\newcommand{\mevc}{\mathrm{MeV}/c}
\newcommand{\mevm}{\mathrm{MeV}/c^2}

\newcommand{\gevm}{\mathrm{GeV}/c^2}

\begin{document}

\preprint{APS/123-QED}

\title{Search for transitions from $\Upsilon(4S)$ and $\Upsilon(5S)$ to $\eta_b(1S)$ and $\eta_b(2S)$ with emission of an $\omega$ meson}

\noaffiliation
\affiliation{University of the Basque Country UPV/EHU, 48080 Bilbao}
\affiliation{University of Bonn, 53115 Bonn}
\affiliation{Brookhaven National Laboratory, Upton, New York 11973}
\affiliation{Budker Institute of Nuclear Physics SB RAS, Novosibirsk 630090}
\affiliation{Faculty of Mathematics and Physics, Charles University, 121 16 Prague}
\affiliation{Chonnam National University, Gwangju 61186}
\affiliation{University of Cincinnati, Cincinnati, Ohio 45221}
\affiliation{Deutsches Elektronen--Synchrotron, 22607 Hamburg}
\affiliation{University of Florida, Gainesville, Florida 32611}
\affiliation{Key Laboratory of Nuclear Physics and Ion-beam Application (MOE) and Institute of Modern Physics, Fudan University, Shanghai 200443}
\affiliation{Justus-Liebig-Universit\"at Gie\ss{}en, 35392 Gie\ss{}en}
\affiliation{SOKENDAI (The Graduate University for Advanced Studies), Hayama 240-0193}
\affiliation{Gyeongsang National University, Jinju 52828}
\affiliation{Department of Physics and Institute of Natural Sciences, Hanyang University, Seoul 04763}
\affiliation{University of Hawaii, Honolulu, Hawaii 96822}
\affiliation{High Energy Accelerator Research Organization (KEK), Tsukuba 305-0801}
\affiliation{J-PARC Branch, KEK Theory Center, High Energy Accelerator Research Organization (KEK), Tsukuba 305-0801}
\affiliation{Higher School of Economics (HSE), Moscow 101000}
\affiliation{Forschungszentrum J\"{u}lich, 52425 J\"{u}lich}
\affiliation{Hiroshima Institute of Technology, Hiroshima 731-5193}
\affiliation{IKERBASQUE, Basque Foundation for Science, 48013 Bilbao}
\affiliation{Indian Institute of Science Education and Research Mohali, SAS Nagar, 140306}
\affiliation{Indian Institute of Technology Guwahati, Assam 781039}
\affiliation{Indian Institute of Technology Hyderabad, Telangana 502285}
\affiliation{Indian Institute of Technology Madras, Chennai 600036}
\affiliation{Indiana University, Bloomington, Indiana 47408}
\affiliation{Institute of High Energy Physics, Chinese Academy of Sciences, Beijing 100049}
\affiliation{Institute of High Energy Physics, Vienna 1050}
\affiliation{Institute for High Energy Physics, Protvino 142281}
\affiliation{INFN - Sezione di Napoli, 80126 Napoli}
\affiliation{INFN - Sezione di Torino, 10125 Torino}
\affiliation{Advanced Science Research Center, Japan Atomic Energy Agency, Naka 319-1195}
\affiliation{J. Stefan Institute, 1000 Ljubljana}
\affiliation{Institut f\"ur Experimentelle Teilchenphysik, Karlsruher Institut f\"ur Technologie, 76131 Karlsruhe}
\affiliation{Kavli Institute for the Physics and Mathematics of the Universe (WPI), University of Tokyo, Kashiwa 277-8583}
\affiliation{Kennesaw State University, Kennesaw, Georgia 30144}
\affiliation{Korea Institute of Science and Technology Information, Daejeon 34141}
\affiliation{Korea University, Seoul 02841}
\affiliation{Kyoto Sangyo University, Kyoto 603-8555}
\affiliation{Kyungpook National University, Daegu 41566}
\affiliation{Universit\'{e} Paris-Saclay, CNRS/IN2P3, IJCLab, 91405 Orsay}
\affiliation{P.N. Lebedev Physical Institute of the Russian Academy of Sciences, Moscow 119991}
\affiliation{Liaoning Normal University, Dalian 116029}
\affiliation{Faculty of Mathematics and Physics, University of Ljubljana, 1000 Ljubljana}
\affiliation{Ludwig Maximilians University, 80539 Munich}
\affiliation{Luther College, Decorah, Iowa 52101}
\affiliation{University of Maribor, 2000 Maribor}
\affiliation{Max-Planck-Institut f\"ur Physik, 80805 M\"unchen}
\affiliation{School of Physics, University of Melbourne, Victoria 3010}
\affiliation{University of Mississippi, University, Mississippi 38677}
\affiliation{University of Miyazaki, Miyazaki 889-2192}
\affiliation{Moscow Physical Engineering Institute, Moscow 115409}
\affiliation{Graduate School of Science, Nagoya University, Nagoya 464-8602}
\affiliation{Kobayashi-Maskawa Institute, Nagoya University, Nagoya 464-8602}
\affiliation{Universit\`{a} di Napoli Federico II, 80126 Napoli}
\affiliation{Nara Women's University, Nara 630-8506}
\affiliation{National Central University, Chung-li 32054}
\affiliation{National United University, Miao Li 36003}
\affiliation{Department of Physics, National Taiwan University, Taipei 10617}
\affiliation{H. Niewodniczanski Institute of Nuclear Physics, Krakow 31-342}
\affiliation{Nippon Dental University, Niigata 951-8580}
\affiliation{Niigata University, Niigata 950-2181}
\affiliation{Novosibirsk State University, Novosibirsk 630090}
\affiliation{Osaka City University, Osaka 558-8585}
\affiliation{Pacific Northwest National Laboratory, Richland, Washington 99352}
\affiliation{Panjab University, Chandigarh 160014}
\affiliation{University of Pittsburgh, Pittsburgh, Pennsylvania 15260}
\affiliation{Punjab Agricultural University, Ludhiana 141004}
\affiliation{Research Center for Nuclear Physics, Osaka University, Osaka 567-0047}
\affiliation{Meson Science Laboratory, Cluster for Pioneering Research, RIKEN, Saitama 351-0198}
\affiliation{Department of Modern Physics and State Key Laboratory of Particle Detection and Electronics, University of Science and Technology of China, Hefei 230026}
\affiliation{Seoul National University, Seoul 08826}
\affiliation{Showa Pharmaceutical University, Tokyo 194-8543}
\affiliation{Soochow University, Suzhou 215006}
\affiliation{Soongsil University, Seoul 06978}
\affiliation{Sungkyunkwan University, Suwon 16419}
\affiliation{School of Physics, University of Sydney, New South Wales 2006}
\affiliation{Department of Physics, Faculty of Science, University of Tabuk, Tabuk 71451}
\affiliation{Tata Institute of Fundamental Research, Mumbai 400005}
\affiliation{Department of Physics, Technische Universit\"at M\"unchen, 85748 Garching}
\affiliation{School of Physics and Astronomy, Tel Aviv University, Tel Aviv 69978}
\affiliation{Toho University, Funabashi 274-8510}
\affiliation{Department of Physics, Tohoku University, Sendai 980-8578}
\affiliation{Earthquake Research Institute, University of Tokyo, Tokyo 113-0032}
\affiliation{Department of Physics, University of Tokyo, Tokyo 113-0033}
\affiliation{Utkal University, Bhubaneswar 751004}
\affiliation{Virginia Polytechnic Institute and State University, Blacksburg, Virginia 24061}
\affiliation{Wayne State University, Detroit, Michigan 48202}
\affiliation{Yamagata University, Yamagata 990-8560}
\affiliation{Yonsei University, Seoul 03722}

\author{P.~Oskin}\affiliation{P.N. Lebedev Physical Institute of the Russian Academy of Sciences, Moscow 119991} 
\author{R.~Mizuk}\affiliation{P.N. Lebedev Physical Institute of the Russian Academy of Sciences, Moscow 119991}\affiliation{Higher School of Economics (HSE), Moscow 101000} 
  \author{H.~Aihara}\affiliation{Department of Physics, University of Tokyo, Tokyo 113-0033} 
  \author{D.~M.~Asner}\affiliation{Brookhaven National Laboratory, Upton, New York 11973} 
  \author{H.~Atmacan}\affiliation{University of Cincinnati, Cincinnati, Ohio 45221} 
  \author{V.~Aulchenko}\affiliation{Budker Institute of Nuclear Physics SB RAS, Novosibirsk 630090}\affiliation{Novosibirsk State University, Novosibirsk 630090} 
  \author{T.~Aushev}\affiliation{Higher School of Economics (HSE), Moscow 101000} 
  \author{R.~Ayad}\affiliation{Department of Physics, Faculty of Science, University of Tabuk, Tabuk 71451} 
  \author{P.~Behera}\affiliation{Indian Institute of Technology Madras, Chennai 600036} 
  \author{K.~Belous}\affiliation{Institute for High Energy Physics, Protvino 142281} 
% ...
\collaboration{The Belle Collaboration}

\begin{abstract}
Using data collected in the Belle experiment at the KEKB asymmetric-energy $e^+e^-$ collider we search for transitions $\Upsilon(4S) \rightarrow \eta_b(1S)\omega$, $\Upsilon(5S) \rightarrow \eta_b(1S)\omega$ and $\Upsilon(5S) \rightarrow \eta_b(2S)\omega$. No significant signals are observed and we set 90\% confidence level upper limits on the corresponding visible cross sections: {$0.2 ~\textrm{pb}, 0.4 ~\textrm{pb}$ and $1.9 ~\textrm{pb}$}, respectively.
\end{abstract}


\maketitle
\section{Introduction}
Recently Belle observed the $\Upsilon(4S) \rightarrow h_b(1P)\eta$ transition and measured its branching fraction to be $\mathcal{B}[\Upsilon(4S) \rightarrow h_b(1P)\eta] =  (2.18 \pm 0.21) \times 10^{-3}$  \cite{Tamponi:2015xzb}. This value is unexpectedly large in comparison with  branching fractions of the $\Upsilon(4S) \rightarrow \Upsilon(1S,2S) \pi^{+}\pi^{-}$ decays \cite{Guido:2017cts,Adachi:2011ji,Aubert:2008az}, and {represents a} strong violation of the Heavy Quark Spin Symmetry (HQSS) \cite{Bondar:2016hva}. A possible mechanism for the HQSS breaking is an admixture of $B\Bar{B}$ pairs in the $\Upsilon(4S)$ state~\cite{Voloshin:2012dk}. Indeed, the $B\Bar{B}$ pair is not an eigenstate of the $b$ quark spin and contains $b\bar{b}$ quarks in both spin-triplet and spin-singlet states. The $\Upsilon(4S)\rightarrow h_b(1P)\eta$ transition proceeds via the spin-singlet component. The transition $\Upsilon(4S)\rightarrow \eta_b(1S)\omega$ might also proceed via the spin-singlet component and thus could be enhanced~\cite{Voloshin:2012dk}. Here we perform a search for the $\Upsilon(4S) \rightarrow \eta_b(1S)\omega$ as well as $\Upsilon(10860) \rightarrow \eta_b(1S)\omega$ and $\Upsilon(10860) \rightarrow \eta_b(2S)\omega$ transitions. {For brevity, the $\Upsilon(10860)$ state is denoted as $\Upsilon(5S)$ according to its quark model assignment.}

We use {the} full data samples of 711 $\textrm f \textrm b^{-1}$ and 121 $\textrm f\textrm b^{-1}$ collected at the $\Upsilon(4S)$ and $\Upsilon(5S)$ resonances by the Belle detector \cite{Abashian:2000cg} at the KEKB asymmetric-energy $e^+e^-$ collider \cite{KEK-B:2002}. The average center-of-mass (c.m.) energy of the $\Upsilon(10860)$ sample is $\sqrt{s}=10.867\,$GeV. The Belle detector is a large-solid-angle magnetic spectrometer that consists of a silicon vertex detector, a 50-layer central drift chamber, an array of aerogel threshold Cherenkov counters, a barrel-like arrangement of time-of-flight scintillation counters, and an electromagnetic calorimeter comprised of CsI(Tl) crystals (ECL) located inside a superconducting solenoid coil that provides a 1.5\,T magnetic field. An iron flux-return located outside of the coil is instrumented to detect $K^{0}_{L}$ mesons and to identify muons.

{For the Monte Carlo (MC) simulation, we use EvtGen \cite{Lange:2001uf} VectorISR model, which correctly describes the angular distribution of ISR photons but uses a flat distribution in photon-energy radiator function. We apply corrections on the ISR photon energy according to Ref. \cite{Benayoun:1999hm}.}  We use the GEANT3 \cite{geant3} package to simulate the detector response.

\section{Event selection}
Since the $\eta_b(1S,2S)$ mesons do not have decay channels that
are convenient to reconstruct, we reconstruct only $\omega\rightarrow\pi^+\pi^-\pi^0$ and use the recoil mass \( M_\textrm{recoil}(\omega) = \sqrt{(\sqrt{s}-E^*_{\omega})^2-(p^*_{\omega})^2} \) to identify the signal, where $E^*_{\omega}$ and $p^*_{\omega}$ are the energy and momentum of the $\omega$ meson in the {c.m.} frame. 

We use {a} generic hadronic event selection with requirements on the position of the primary vertex, track multiplicity, and the total energy and momentum of the event \cite{Abe:2001hj}. For charged pions we require the distance of closest approach to the interaction point to be within 2\,cm along the beam direction and within 0.2\,cm in the plane transverse to the beam direction. We apply loose particle identification requirements to separate charged pions from kaons, protons and electrons. The energy of {a photon} in the laboratory frame is required to be greater than $50\,\mev$ for the barrel part of {the} ECL and greater than $100\,\mev$ for the endcap part of {the} ECL. To further suppress the background from low-energy photons, we require the momentum of the $\pi^{0}$ in the c.m.\ frame to be above 240$\,\mevc$, 270$\,\mevc$ and $140\,\mevc$ for the $\Upsilon(4S) \rightarrow \eta_b(1S)\omega$, $\Upsilon(5S) \rightarrow \eta_b(1S)\omega$ and $\Upsilon(5S) \rightarrow \eta_b(2S)\omega$ transitions, respectively. The masses of the $\pi^{0}$ and $\omega$ candidates should satisfy $|M(\gamma\gamma)-m_{\pi^0}|<8\,\mevm$ and $|M(\pi^+\pi^-\pi^0)-m_\omega|<12\,\mevm$ \cite{PDG2020}. The resolutions in $M(\gamma\gamma)$ and $M(\pi^+\pi^-\pi^0)$ are $5.5\,\mevm$ and $8\,\mevm$, respectively.

The $\omega\rightarrow\pi^+\pi^-\pi^0$ {events predominantly populate the central part of the Dalitz Plot (DP), while background events populate the region near the boundaries.} Therefore we require the normalized distance from the DP {center}, $r$, to be lower than $0.84$. The variable $r$ takes values from $r$=0 at the DP {center} to $r$=1 at its boundary~\cite{Matvienko:2015gqa}.

To suppress background from continuum events $e^{+}e^{-} \rightarrow q\Bar{q}$  ($q = u,d,s,c$) that have a jet-like shape, we use the angle $\theta_\textrm{thrust}$ between the thrust axis of $\pi^+\pi^-\pi^0$ and the thrust axis of the rest of the event in the $\eta_{b}$-meson candidate rest frame. The thrust axis is defined as the unit vector $\vec{n}_{T}$, which maximizes the thrust value: $T = \sum_{i} |\vec{p}_{i} \cdot \vec{n}_T|/ \sum_{i} |\vec{p}_{i}|$. The selection criteria are $|\cos(\theta_\textrm{thrust})|<0.90$ and $|\cos(\theta_\textrm{thrust})|<0.70$ for the $\Upsilon(4S) \rightarrow \eta_b(1S)\omega$ and $\Upsilon(5S) \rightarrow \eta_b(1S)\omega$ transitions, respectively, while for the $\Upsilon(5S) \rightarrow \eta_b(2S)\omega$ transition no criteria on $\cos(\theta_\textrm{thrust})$ are applied.

The selection requirements {described above} are optimized using a figure of merit $N_\textrm{sig}/\sqrt{N_\textrm{bkg}}$, where the signal yield is estimated using MC simulation, and the background yield is estimated using the signal region in data, assuming the signal fraction is very small. For optimization of various requirements we use an iterative procedure. 
The $M_\textrm{recoil}(\pi^{+}\pi^{-}\pi^{0})$ is required to be in the interval $(9.20, 9.60)\,\gevm$ for the $\eta_b(1S)$ candidates and $(9.90, 10.05)\,\gevm$ for the $\eta_b(2S)$ candidates. These intervals are used in the fits described below.

\begin{figure}[htbp]
\begin{center}
\includegraphics[height=6cm, keepaspectratio]{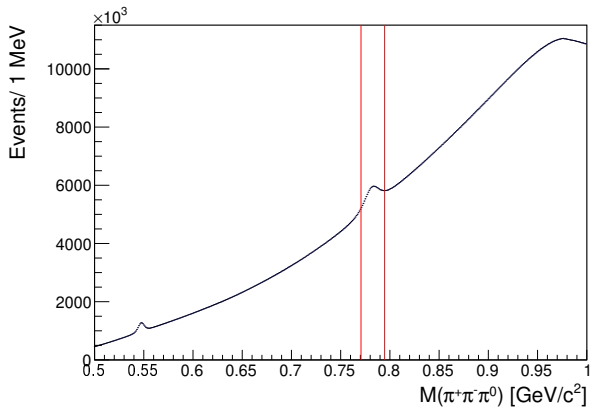}
\caption{\label{fig:pipipi}The $M(\pi^+\pi^-\pi^0)$ distribution for the $\Upsilon(4S) \rightarrow \eta_b(1S)\omega$ candidates. Red lines indicate the $\omega$ mass requirement.}
\end{center}
\end{figure}

The $M(\pi^+\pi^-\pi^0)$ distribution for the $\Upsilon(4S) \rightarrow \eta_b(1S)\omega$ candidates without the $\omega$ mass requirement is shown in Fig \ref{fig:pipipi}. One can see clear signals of the $\eta$ and $\omega$ mesons.
{The purity of the $\omega$-meson signal is estimated to be 13\%. In
the $\Upsilon(5S)\to\eta_b(1S)\omega$ and $\Upsilon(5S)\to\eta_b(2S)\omega$ transitions the purities are 24\% and 6\%, respectively.}

{We find on average 1.33, 1.12 and 1.63 candidates per event for the $\Upsilon(4S) \rightarrow \eta_b(1S)\omega$, $\Upsilon(5S) \rightarrow \eta_b(1S)\omega$ and $\Upsilon(5S) \rightarrow \eta_b(2S)\omega$ transitions, respectively. The multiple candidates are not correlated in $M_\textrm{recoil}(\pi^{+}\pi^{-}\pi^{0})$, therefore we do not perform a best candidate selection.} The total {selection} efficiencies are {5.5\%, 5.6\% and 6.6\%} for the three transitions, respectively.

\section{Fit to data}

We perform a binned $\chi^2$ fit to the $M_\textrm{recoil}(\pi^{+}\pi^{-}\pi^{0})$ distributions in the $\eta_b(1S)$ and $\eta_b(2S)$ mass regions. The fit function is a sum of signal and background components. The signal component is described with a two-sided Crystal Ball (CB) function, which consists of a Gaussian core portion and power-law tails; the function and its first derivative are both continuous \cite{Skwarnicki:1986xj}. The parameters of the CB function are fixed using MC simulation; the $\sigma$-parameters of the core Gaussian are $12.7\,\mevm$, {$14.6\,\mevm$ and $9.2\,\mevm$} for $\Upsilon(4S) \rightarrow \eta_b(1S)\omega$, $\Upsilon(5S) \rightarrow \eta_b(1S)\omega$ and $\Upsilon(5S) \rightarrow \eta_b(2S)\omega$, respectively. The integral of the signal function over the fit range is taken as a signal yield. The background component is described with a Chebyshev polynomial; its order is chosen as the one that gives the
maximum $p$-value {for} the fit. The polynomial orders are 8, 5 and {6} for the three transitions, respectively. The $M_\textrm{recoil}(\pi^{+}\pi^{-}\pi^{0})$ distributions and fit results for $\Upsilon(4S) \rightarrow \eta_b(1S)\omega$, $\Upsilon(5S) \rightarrow \eta_b(1S)\omega$ and $\Upsilon(5S) \rightarrow \eta_b(2S)\omega$ are shown in Figs. 2, 3 and 4, respectively. We use 1 MeV {bins} for fitting and 10 MeV {bins} for visualization to improve clarity. No significant signals are observed. The obtained signal yields for each transition are presented in Table \ref{table:sigma_values}.

To set upper limits on the branching fractions, we study the systematic uncertainties in the yields. We vary the $\eta_b(1S)$ and $\eta_b(2S)$ masses and widths within one standard deviation \cite{PDG2020}. The $\eta_b(2S)$ width is estimated using a model-independent relation \cite{Voloshin:2007dx}:

\begin{equation}
\begin{split}
\Gamma[\eta_b(2S)] &= \Gamma[\eta_b(1S)]\cdot \frac{\Gamma[\Upsilon(2S)\rightarrow e^{+}e^{-}]}{\Gamma[\Upsilon(1S)\rightarrow e^{+}e^{-}]} \\
&= 4.6^{+2.3}_{-1.8} ~ \textrm{MeV}.
\end{split}
\end{equation}

\begin{figure}[htbp]
\includegraphics[height=8.5cm, keepaspectratio]{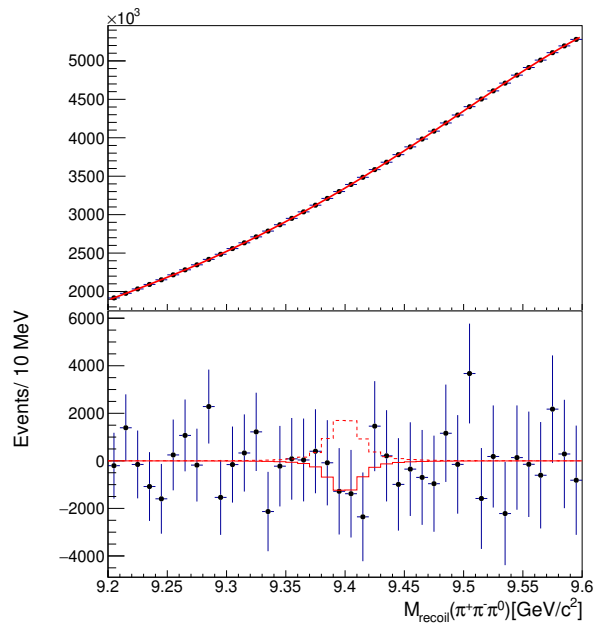}
\caption{\label{fig:y4s_data_fit}The $M_\textrm{recoil}(\pi^{+}\pi^{-}\pi^{0})$ distribution for the $\Upsilon(4S) \rightarrow \eta_b(1S)\omega$ candidates.
Top: data points with the fit function overlaid; {note the suppressed zero on the vertical scale.}
Bottom: {residuals between the data and the fit function. The solid line shows the fit function for the best fit; the dashed line shows the same function with the yield fixed to the upper limit.}}
\end{figure}

\begin{figure}[htbp]
\includegraphics[height=8.5cm, keepaspectratio]{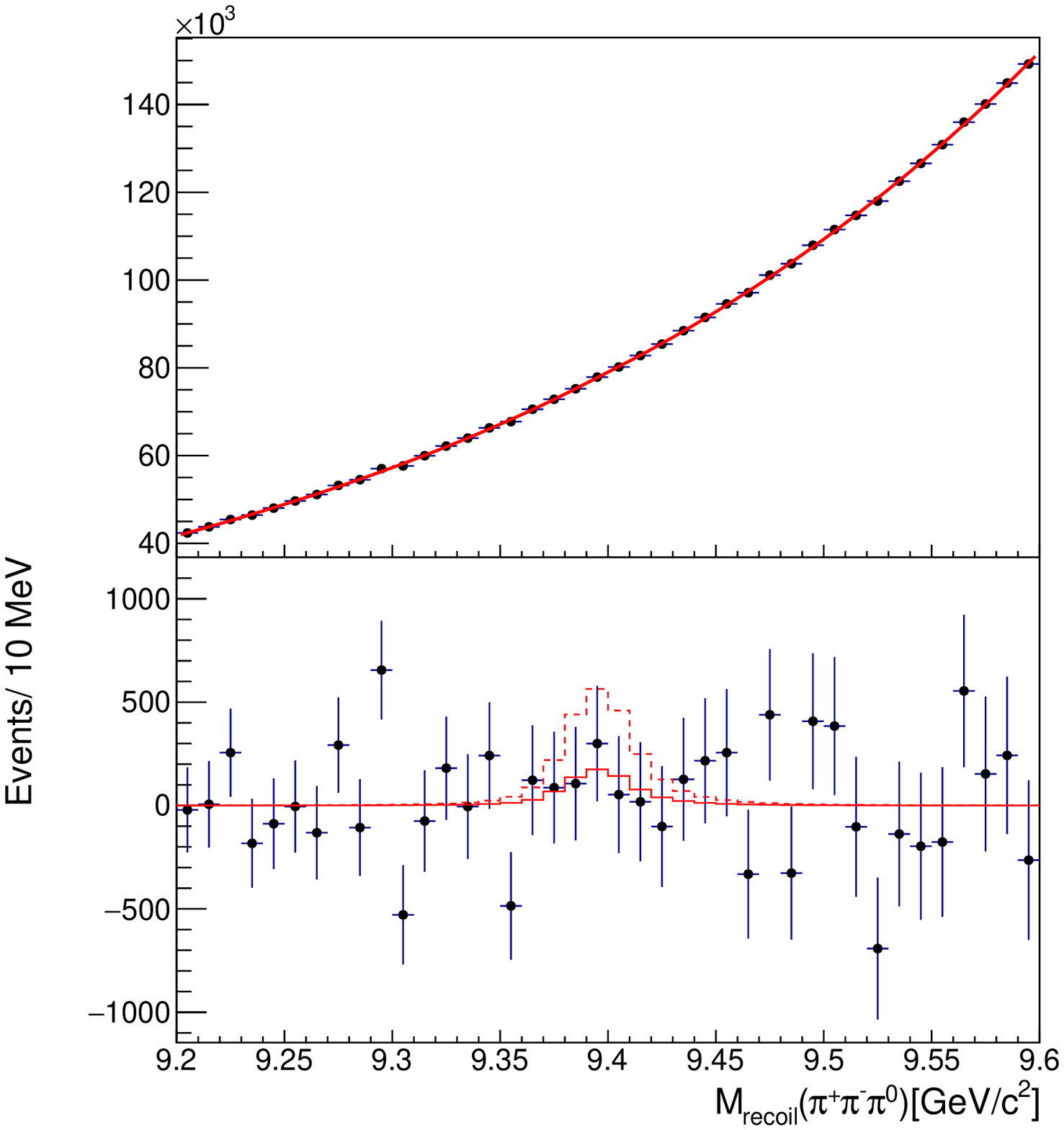}
\caption{\label{fig:y5s1_data_fit}The $M_\textrm{recoil}(\pi^{+}\pi^{-}\pi^{0})$ distribution for the $\Upsilon(5S) \rightarrow \eta_b(1S)\omega$ candidates; {note the suppressed zero on the vertical scale. Symbols are the same as those in Fig. \ref{fig:y4s_data_fit}.}}
\end{figure}

\begin{figure}[htbp]
\includegraphics[height=8.5cm, keepaspectratio]{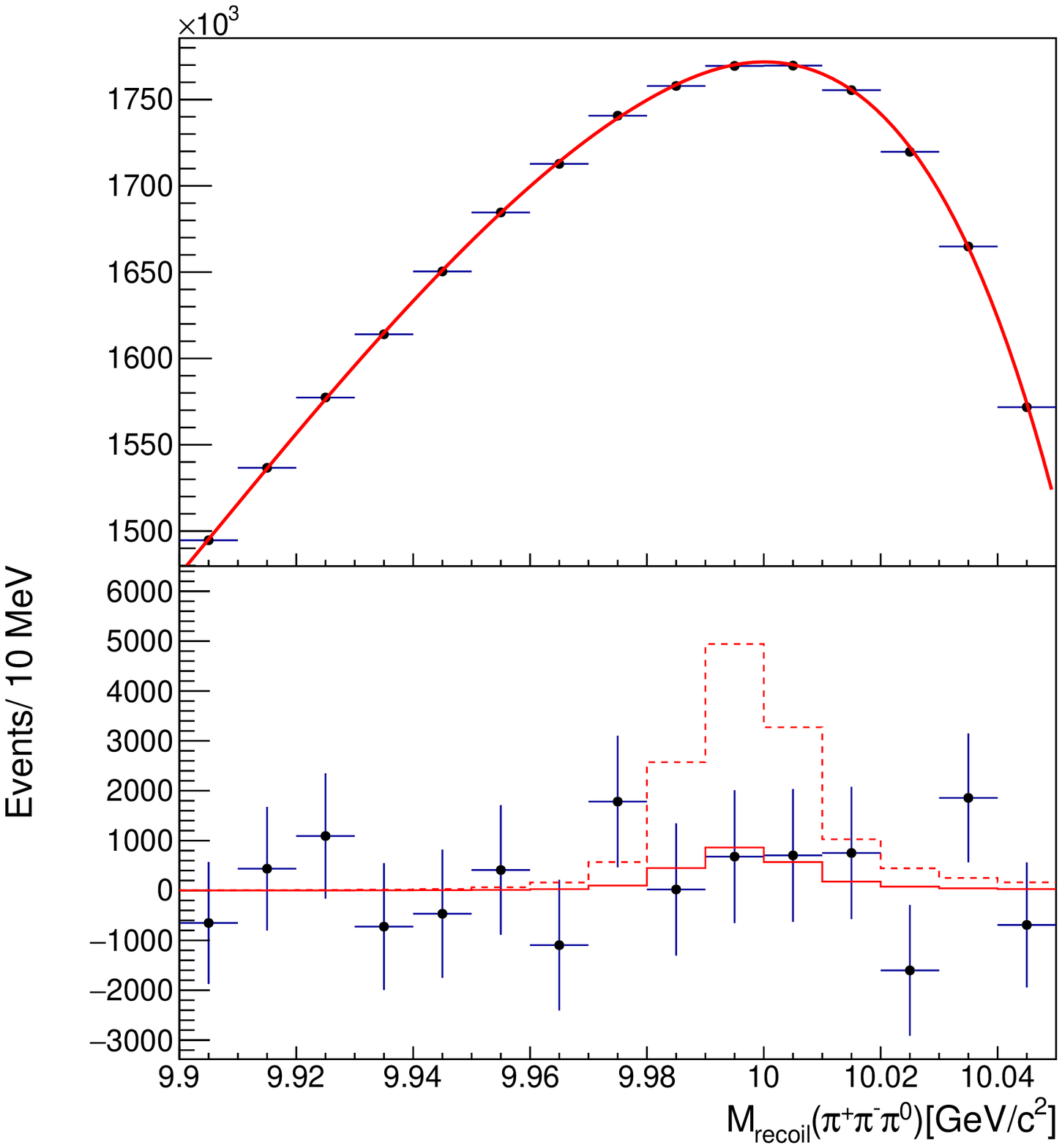}
\caption{\label{fig:y5s2_data_fit}The $M_\textrm{recoil}(\pi^{+}\pi^{-}\pi^{0})$ distribution for the $\Upsilon(5S) \rightarrow \eta_b(2S)\omega$ candidates; {note the suppressed zero on the vertical scale. Symbols are the same as those in Fig. \ref{fig:y4s_data_fit}.}}
\end{figure}

\renewcommand{\arraystretch}{1.3}
\begin{table}[htbp]
\caption{Signal yields for various transitions obtained from the fits in units of $10^3$.}
\begin{ruledtabular}
\begin{tabular}{ccc}
$\Upsilon(4S) \rightarrow \eta_b(1S)\omega$ & $\Upsilon(5S) \rightarrow \eta_b(1S)\omega$ & $\Upsilon(5S) \rightarrow \eta_b(2S)\omega$       \\
\hline
${-5.0 \pm 6.6^{+1.5}_{-2.3}}$  & ${0.8 \pm 1.0^{+0.3}_{-0.1}}$  &  ${2.4 \pm 5.8^{+\phantom{0}3.2}_{-10.8}}$  \\
\end{tabular}
\label{table:sigma_values}
\end{ruledtabular}
\end{table}
\renewcommand{\arraystretch}{1}

\noindent {The ISR tails are sensitive to the energy dependence of the $e^+e^-\to\eta_b(nS)\omega$ cross sections. Instead of a resonant
production via $\Upsilon(4S)$ or $\Upsilon(5S)$, we consider also the cross sections that are energy independent. We generate MC samples for each modification and use them to determine signal shapes.} To estimate the uncertainties due to background parameterization we vary the fit range and increase or decrease the polynomial order by one. Maximal deviations in each case are considered to be a systematic uncertainty. The summary of the uncertainties in the yields is presented in Table \ref{table:total_fit_err}. The total systematic uncertainty in the yield is found by adding various contributions in quadrature.

\renewcommand{\arraystretch}{1.3}
\begin{table}[htbp]
\caption{Systematic uncertainties in the yields of various transitions in units of $10^{2}$.}
\begin{ruledtabular}
\begin{tabular}{l|ccc}
      & $\Upsilon(4S) \rightarrow$    & \multicolumn{2}{c}{$\Upsilon(5S) \rightarrow$}  \\
      & $\eta_b(1S)\omega$ & $\eta_b(1S)\omega$ & $\eta_b(2S)\omega$          \\
\hline
$\eta_b$ mass & ${^{+6}_{-3}}$  & ${\pm 0}$ & ${^{+27}_{-34}}$ \\
$\eta_b$ width & ${^{+6}_{-7}}$ & ${\pm 1}$ & ${^{+5}_{-4}}$ \\
ISR tail & ${^{+\phantom{0}0}_{-12}}$  & ${^{+2}_{-0}}$ & ${^{+2}_{-0}}$ \\
Background & ${^{+11}_{-17}}$  & ${^{+3}_{-1}}$ & ${^{+\phantom{0}17}_{-102}}$ \\
\hline
Total &${^{+15}_{-23}}$  & ${^{+3}_{-1}}$ & ${^{+\phantom{0}32}_{-108}}$
\end{tabular}
\label{table:total_fit_err}
\end{ruledtabular}
\end{table}
\renewcommand{\arraystretch}{1}

\section{Upper limits on visible cross sections and branching fractions}

{The visible cross sections are calculated as:}

\begin{equation}
{\sigma_\textrm{vis}[e^+e^- \rightarrow \eta_b(mS)\omega] =} {\frac{N}{\epsilon \cdot \mathcal{B}[\omega \rightarrow \pi^{+}\pi^{-}\pi^{0}] \cdot L_{\Upsilon(nS)}}},
\end{equation}

where $n$=4, 5 and $m$=1, 2; $N$ is the signal yield, $\epsilon$ is the selection efficiency, $L_{\Upsilon(nS)}$ is the integrated luminosity, $L_{\Upsilon(4S)}=711~\textrm{fb}^{-1}$ and $L_{\Upsilon(5S)}=121.4~\textrm{fb}^{-1}$.

We take into account the uncertainty in the efficiency due to possible {discrepancies} between data and MC simulation (1\% per track and 2.2\% for $\pi^0$), the uncertainty in $\mathcal{B}[\omega \rightarrow \pi^{+} \pi^{-} \pi^{0}]$ \cite{PDG2020} and the uncertainty in the $\Upsilon(4S)$ and $\Upsilon(5S)$ integrated luminosity. The total multiplicative uncertainty in the visible cross section is 4.5\%.

To combine the uncertainty in the yield, $\delta_N$, which is obtained by adding corresponding statistical and systematic uncertainties in quadrature, and the multiplicative uncertainty $\delta$, we use formula:

\begin{equation}
(N \pm \delta_N) \cdot (1 \pm \delta) = N \pm (\delta_N \oplus N\delta \oplus \delta_N\delta),
\end{equation}
where the symbol $\oplus$ denotes addition in quadrature.

Estimated {visible cross sections} and upper limits set using the Feldman-Cousins method \cite{Feldman:1997qc} at the 90\% confidence level are presented in Table \ref{table:CS_data}.

\renewcommand{\arraystretch}{1.3}
\begin{table}[htbp]
\caption{{Visible cross sections and upper limits at the 90\% confidence level in pb.}}
\begin{ruledtabular}
\begin{tabular}{c|cc}
      & Visible cross section & Upper limit  \\
\hline
$e^+e^-\to\eta_b(1S)\omega$ at $\Upsilon(4S)$ & ${-0.14 ^{+0.19}_{-0.20}}$ & ${0.2}$ \\
$e^+e^-\to\eta_b(1S)\omega$ at $\Upsilon(5S)$& ${\phantom{-}0.13 ^{+0.18}_{-0.17}}$ & ${0.4}$ \\
$e^+e^-\to\eta_b(2S)\omega$ at $\Upsilon(5S)$& ${0.3 ^{+0.9}_{-1.7}}$ & ${1.9}$
\end{tabular}
\label{table:CS_data}
\end{ruledtabular}
\end{table}

{We also estimate branching fractions of the $\Upsilon(4S) \rightarrow \eta_b(1S)\omega$, $\Upsilon(5S) \rightarrow \eta_b(1S)\omega$ and $\Upsilon(5S) \rightarrow \eta_b(2S)\omega$ transitions using the number of $\Upsilon(4S)$ or $\Upsilon(5S)$ instead of the luminosity in the denominator of Eq.(2). The number of $\Upsilon(4S)$ is $(771.6\pm10.6)\times10^6$, while the number of $\Upsilon(5S)$ is estimated as $L_{\Upsilon(5S)}\cdot\sigma_{b\bar{b}}$, where $\sigma_{b\bar{b}}=(0.340\pm0.016)\,\mathrm{nb}$~\cite{Tamponi:2018cuf}. The total multiplicative uncertainty in the branching fraction is 4.5\% and 6.5\% for transitions from $\Upsilon(4S)$ and $\Upsilon(5S)$, respectively. Estimated branching fractions and upper limits at the 90\% confidence level are presented in Table \ref{table:BF_data}.}

\begin{table}[htbp]
\caption{Branching fractions and upper limits at the 90\% confidence level.}
\begin{ruledtabular}
\begin{tabular}{c|cc}
      & Branching fraction & Upper limit  \\ 
\hline
$\Upsilon(4S) \rightarrow \eta_b(1S)\omega$ & ${(-1.3 ^{+1.8}_{-1.9})\times 10^{-4}}$ & ${1.8 \times 10^{-4}}$ \\
$\Upsilon(5S) \rightarrow \eta_b(1S)\omega$ & ${(\phantom{-}3.7 ^{+5.4}_{-5.1}) \times 10^{-4}}$ & ${1.3 \times 10^{-3}}$ \\
$\Upsilon(5S) \rightarrow \eta_b(2S)\omega$ & ${(\phantom{-}1.0 ^{+2.7}_{-5.1}) \times 10^{-3}}$ & ${5.6 \times 10^{-3}}$
\end{tabular}
\label{table:BF_data}
\end{ruledtabular}
\end{table}

\renewcommand{\arraystretch}{1}

{We also set the upper limit on the ratio:}
\begin{equation}
{\frac{\mathcal{B} [\Upsilon(4S) \longrightarrow \eta_b(1S)\omega]}{\mathcal{B} [\Upsilon(4S) \rightarrow h_b(1P)\eta]} < 8.4 \times 10^{-2}}
\end{equation}
{at the 90\% confidence level.}

\section{Crosscheck with $\Upsilon(5S) \rightarrow \chi_{bJ}(1P)\omega$}
\begin{figure}[htbp]
\includegraphics[height=8.5cm, keepaspectratio]{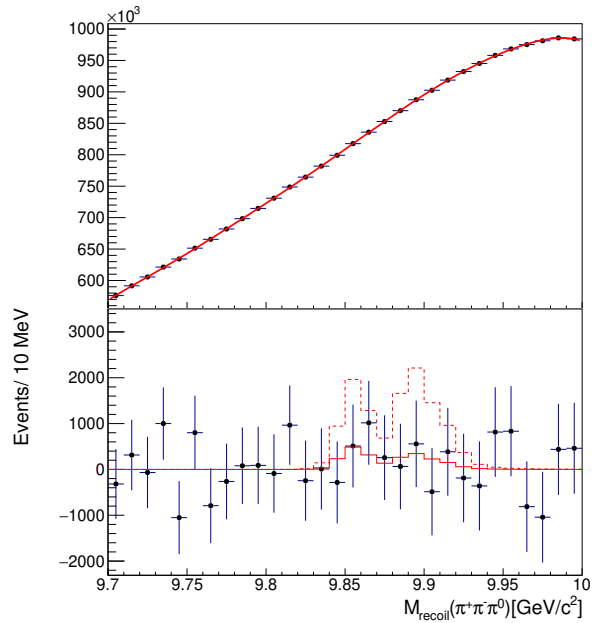}
\caption{\label{fig:chibJ_data_fit} The $M_\textrm{recoil}(\pi^{+}\pi^{-}\pi^{0})$ distribution for the $\Upsilon(5S) \rightarrow \chi_{bJ}(1P)\omega$ candidates; {note the suppressed zero on the vertical scale. Symbols are the same as those in Fig. \ref{fig:y4s_data_fit}.}}
\end{figure}

As a crosscheck, we perform a search for the $\Upsilon(5S) \rightarrow \chi_{bJ}(1P)\omega$ transitions that were observed previously using exclusive reconstruction \cite{He:2014sqj}. The analysis procedure is the same as for the $\eta_b(nS)\omega$ transitions. To fit the signal region we use the sum of three CB functions corresponding to $\chi_{b0}(1P), \chi_{b1}(1P)$ and $ \chi_{b2}(1P)$ signals. Since we do not have enough resolution to measure $\chi_{b1}(1P)$ and $\chi_{b2}(1P)$ yields individually, we fix the ratio between them according to the known values \cite{He:2014sqj}. To fit the background contribution we use a ninth-order Chebyshev polynomial. The fit result is shown in Fig. \ref{fig:chibJ_data_fit}.

\noindent There are no significant signals. The obtained upper limits on branching fractions at the 90\% confidence level are

\begin{equation}
\begin{split}
\mathcal{B}[\Upsilon(5S) \rightarrow \chi_{b0}(1P)\omega] <  {3.0 \times 10^{-3}}, \\
\mathcal{B}[\Upsilon(5S) \rightarrow \chi_{b1}(1P)\omega] <  {3.2 \times 10^{-3}}. 
\end{split}
\end{equation} 

\noindent Since the ratio between $\chi_{b1}(1P)$ and $\chi_{b2}(1P)$ is fixed, we do not give {an} upper limit for $\chi_{b2}(1P)$. The obtained upper limits are consistent with the exclusive measurement \cite{He:2014sqj}:

\begin{equation}
\begin{split}
\mathcal{B}[\Upsilon(5S) \rightarrow \chi_{b0}(1P)\omega] &<  3.9 \times 10^{-3}, \\
\mathcal{B}[\Upsilon(5S) \rightarrow \chi_{b1}(1P)\omega] &=  (1.57 \pm 0.30) \times 10^{-3}.  
\end{split}
\end{equation}

\section{Conclusion}
In summary, we perform a search for the transitions $\Upsilon(4S) \longrightarrow \eta_b(1S)\omega$, $\Upsilon(5S) \longrightarrow \eta_b(1S)\omega$ and $\Upsilon(5S) \longrightarrow \eta_b(2S)\omega$. No significant signals are observed and we set upper limits on {visible cross sections and branching fractions presented in Tables \ref{table:CS_data} and \ref{table:BF_data}. The upper limit for $\Upsilon(4S) \longrightarrow \eta_b(1S)\omega$ is order of magnitude lower than the value for the similar transition to a spin-singlet state, $\Upsilon(4S) \rightarrow h_b(1P)\eta$~\cite{Tamponi:2015xzb}. We set the upper limit on the ratio:
\begin{equation}
{\frac{\mathcal{B} [\Upsilon(4S) \longrightarrow \eta_b(1S)\omega]}{\mathcal{B} [\Upsilon(4S) \rightarrow h_b(1P)\eta]} < 8.4 \times 10^{-2}}
\end{equation}
at the 90\% confidence level. As both transitions are expected to proceed via the $B\bar{B}$ admixture in the $\Upsilon(4S)$ state~\cite{Voloshin:2012dk}, our result will help to better understand this mechanism.} The suppression of the $\Upsilon(4S)\to\eta_b(1S)\omega$ transition relative to the $\Upsilon(4S)\to h_b(1P)\eta$ one could be due to different overlaps between the initial state and the bottomonium in the final state or the details of the $\eta$ and $\omega$ meson production.

\section{Acknowledgments}
We thank the KEKB group for the excellent operation of the
accelerator; the KEK cryogenics group for the efficient
operation of the solenoid; and the KEK computer group, and the Pacific Northwest National
Laboratory (PNNL) Environmental Molecular Sciences Laboratory (EMSL)
computing group for strong computing support; and the National
Institute of Informatics, and Science Information NETwork 5 (SINET5) for
valuable network support.  We acknowledge support from
the Ministry of Education, Culture, Sports, Science, and
Technology (MEXT) of Japan, the Japan Society for the 
Promotion of Science (JSPS), and the Tau-Lepton Physics 
Research Center of Nagoya University; 
the Australian Research Council including grants
DP180102629, 
DP170102389, 
DP170102204, 
DP150103061, 
FT130100303; 
Austrian Science Fund (FWF);
the National Natural Science Foundation of China under Contracts
No.~11435013,  
No.~11475187,  
No.~11521505,  
No.~11575017,  
No.~11675166,  
No.~11705209;  
Key Research Program of Frontier Sciences, Chinese Academy of Sciences (CAS), Grant No.~QYZDJ-SSW-SLH011; 
the  CAS Center for Excellence in Particle Physics (CCEPP); 
the Shanghai Pujiang Program under Grant No.~18PJ1401000;  
the Ministry of Education, Youth and Sports of the Czech
Republic under Contract No.~LTT17020;
the Carl Zeiss Foundation, the Deutsche Forschungsgemeinschaft, the
Excellence Cluster Universe, and the VolkswagenStiftung;
the Department of Science and Technology of India; 
the Istituto Nazionale di Fisica Nucleare of Italy; 
National Research Foundation (NRF) of Korea Grant
Nos.~2016R1\-D1A1B\-01010135, 2016R1\-D1A1B\-02012900, 2018R1\-A2B\-3003643,
2018R1\-A6A1A\-06024970, 2018R1\-D1A1B\-07047294, 2019K1\-A3A7A\-09033840,
2019R1\-I1A3A\-01058933;
Radiation Science Research Institute, Foreign Large-size Research Facility Application Supporting project, the Global Science Experimental Data Hub Center of the Korea Institute of Science and Technology Information and KREONET/GLORIAD;
the Polish Ministry of Science and Higher Education and 
the National Science Center;
the Russian Science Foundation (RSF), Grant No. 18-12-00226;
University of Tabuk research grants
S-1440-0321, S-0256-1438, and S-0280-1439 (Saudi Arabia);
the Slovenian Research Agency;
Ikerbasque, Basque Foundation for Science, Spain;
the Swiss National Science Foundation; 
the Ministry of Education and the Ministry of Science and Technology of Taiwan;
and the United States Department of Energy and the National Science Foundation.

\end{document}